%
\input mnrass.sty
\pageoffset{-2.5pc}{0pc}

 

\Autonumber  


\pagerange{000--000}
\pubyear{1996}
\volume{000}

\begintopmatter  

\title{The \lq{Tip}\rq\ of the Red Giant Branch as a distance
indicator: results from evolutionary models}

\author{ Maurizio Salaris$^{1,2}$ \& Santi Cassisi$^{3,4}$}

\affiliation{$^1$Institut d'Estudis Espacials de Catalunya,
E-08034, Barcelona, Spain}
\affiliation{$^2$Max-Planck-Institut f\"ur Astrophysik, D-85740,
Garching, Germany - E-Mail: maurizio@MPA-Garching.mpg.de} 
\affiliation{$^3$Universit\'a degli studi de L'Aquila, Dipartimento di Fisica,
Via Vetoio, I-67100, L'Aquila, Italy}
\affiliation{$^4$Osservatorio Astronomico di Collurania, Via M. Maggini,
 I-64100, Teramo, Italy - E-Mail: cassisi@astrte.te.astro.it}

\shortauthor{M.Salaris \& S.Cassisi}
\shorttitle{The Red Giant \lq{Tip}\rq\ luminosity}


\abstract
\tx

New theoretical evaluations of the Red Giant Branch Tip 
(TRGB) luminosity, by adopting the most updated physical inputs in 
computing canonical stellar models, are presented.  
Theoretical relations for the run of the TRGB bolometric and I magnitude 
with the metallicity are provided together with
a comparison of the distance scale based on these relations and the 
RR Lyrae distance scale presented in Cassisi \& Salaris (1996) 
and the Cepheid distance scale adopted by Lee, Freedman \& Madore (1993).
The result of this comparison - performed  by adopting a sample of galactic globular
clusters and a sample of resolved galaxies - 
discloses a satisfactory agreement between theory and observations at the level of
0.1 mag. This occurrence can be regarded as an evidence for the consistency between theoretical
Red Giant, Horizontal Branch stellar models and independent Cepheid observations, and allows
to safely use the provided TRGB luminosity relations as an alternative 
primary distance indicator for galaxies in which the stellar component has been resolved. 


\maketitle  

\section{Introduction}

\tx 
The Cepheid period-luminosity (P-L) relation is the
basis for the calibration of a wide range of secondary distance
indicators applicable at larger distances than the Cepheids
themselves. However, Cepheid observations are restricted only to
Population I systems and to late-type galaxies, while an excellent
alternative primary distance indicator is the tip of the Red Giant
Branch (TRGB); the use of this indicator results particularly
attractive since it is applicable to all morphological types of
galaxies as long as an old stellar population is present.
After the pioneering studies by Baade (1944) and Sandage (1971)
suggesting the potential usefulness of the TRGB as a distance indicator,
the TRGB method for estimating the distances to several nearby galaxies was used
by many authors (see e.g. Lee, Freedman \& Madore 1993, hereinafter LFM93;
Sakai, Madore \& Freedman 1996, Soria et al. 1996, Elson 1996).

The TRGB
method has an underlying physical basis: the tip of the Red Giant
Branch (RGB) marks the helium ignition in the degenerate He core of
low mass stars, and its luminosity depends on the He core mass,
that is remarkably constant for ages larger than a few Gyr.
According to models found in the literature (see e.g. Straniero \& Chieffi
1991, Castellani, Chieffi \& Straniero 1992 and references therein) 
the bolometric luminosity of the TRGB (at a fixed metallicity Z between
0.0001 and 0.02)
varies by at most 0.05 mag for ages ranging from 10 up to 20 Gyr, and
by $\approx 0.1$ mag if the range is stretched down to $\approx$ 2 Gyr
(the exact value depending on the metallicity). LFM93
have shown that taking into account the Yale isochrones
(Green, Demarque \& King 1987) and the photometric data provided by Da Costa \& Armandroff 
(1990, hereinafter DA90) the theoretical and observational I magnitude of TRGB
stars in globular clusters (GCs) is constant within 0.1 mag 
for $-2.2<[Fe/H]<-0.7$, when the corresponding V magnitude varies by 1.3 mag over the
same metallicity range (see their fig. 1), suggesting therefore the use of
the observed I magnitude of TRGB stars as a distance indicator. 

In particular, LFM93
have provided a calibration of this method in a large range of metallicity, and 
in a subsequent paper Madore \& Freedman (1995) undertook a number of computer
simulations and concluded that the TRGB  method can be successfully used to
determine distances accurate to 0.2 mag for galaxies out to 3 Mpc using
ground based telescopes, and out to a factor of four further in
distance using the Hubble Space Telescope.

In the present paper, we want to investigate the possibility to adopt the TRGB
method for determining distances of nearby galaxies, using results from
stellar models computed by adopting the most updated physical inputs.
More in details, this paper is the third one of a series (see Salaris \& Cassisi 1996
and Cassisi \& Salaris 1997, hereinafter Paper I and II respectively)
devoted to the comparison of standard updated RGB stellar models with
observational data. In Paper I we have discussed the calibration of
the effective temperature of RGB stellar models, while in Paper II 
it is shown that updated RGB models reproduce well
the observed V magnitude of GC RGB luminosity
function bumps ($V_{bump}$), which can be used as a diagnostic of the Hydrogen 
stratification in the stellar interior.
In paper II we also presented new Zero Age Horizontal Branch (ZAHB) models,
computed with the same updated input physics as the RGB ones;
the distance scale obtained from these models has been adopted for comparing
theoretical and observed values of $V_{bump}$.

In this paper we will complete the analysis of our updated 
RGB evolutionary models by comparing the observational determinations of the
TRGB luminosities of a sample of GCs and galaxies with  
theory, adopting the RR Lyrae distance scale given in Paper II and
the Cepheid distance scale used by LFM93.
In this way we will assess in an independent way the reliability 
of our theoretical evolutionary models.
We provide new updated prescriptions for the bolometric and I (Cousins) magnitude
of the RGB tip and for the (V-I) color of observed RGB as a
function of the metallicity, based on our theoretical models and on
recent spectroscopical determinations of the chemical composition in GC stars,
to be applied when using the TRGB method as a distance indicator. 

The theoretical RGB models are presented in section 2, while in
section 3 the comparison with observational data of a sample of GCs
and resolved galaxies 
is presented.
Summary and conclusions follow in section 4.

\section{\bf RGB theoretical models}

\tx

The models used in this paper have been already presented
in Paper I and II. 
To summarize, we have computed
canonical evolutionary models of stars with masses of 0.75$M_{\odot}$, 0.80$M_{\odot}$ 
and
0.90$M_{\odot}$ and for metallicities 
Z=0.0001 - 0.0003 - 0.0006 - 0.001 - 0.003 - 0.006.
As for the original Helium abundance ($Y$),
we have adopted Y=0.23 at all metallicities,
according to the results by Buzzoni et al. (1983), which find an
almost constant He abundance in a sample of GCs spanning
approximately the same range of metallicity, and to the most recent evaluations
for the primordial He abundance (Olive, Skillman \& Steigman 1996).

All the models have
been computed adopting the FRANEC evolutionary code (see Chieffi \&
Straniero 1989). The OPAL opacity tables (Rogers \& Iglesias
1992, Iglesias, Rogers \& Wilson 1992) for $T>10000K$ and the
Alexander \& Ferguson (1994) opacities for lower temperatures have been used. 
Both high and low temperature opacity tables have been computed adopting the
solar heavy elements distribution by Grevesse (1991).
The electronic conduction is treated according to Itoh et al. (1983). 
The equation of state (EOS) by Straniero (1988) has been used, supplemented by 
a Saha EOS at lower temperatures, as described by Chieffi \& Straniero (1989).
\table{1}{S}{\bf Table 1. \rm Luminosity and bolometric magnitude of
the TRGB for the adopted metallicities as obtained by updated stellar models.} 
{\halign{%
\rm#\hfil&\hskip10pt\hfil\rm#\hfil&\hskip10pt\hfil\rm\hfil#&\hskip10pt\hfil\rm\hfil#\cr
$Z$ & [M/H] & $\rm \log(L/L_{\odot})^{tip}$ & $\rm {M}^{tip}_{bol}$ \cr 
\noalign{\vskip 10pt}
0.0001 & -2.35 & 3.296~~~  &  -3.490 \cr 
0.0003 & -1.87 & 3.335~~~  &  -3.587 \cr
0.0006 & -1.57 & 3.361~~~  &  -3.653 \cr   
~0.001 & -1.35 & 3.378~~~  &  -3.695 \cr
~0.003 & -0.87 & 3.416~~~  &  -3.790 \cr
~0.006 & -0.57 & 3.438~~~  &  -3.845 \cr}}

By interpolating among the various computed stellar models at each metallicity,
we have obtained the value of the TRGB luminosity corresponding to an age t=15 Gyr.
Concerning this choice for the average age the interested reader is
referred to Paper II and references therein. However it is worth noting that  
for old stellar systems as the ones we are dealing with, the TRGB
luminosity varies only by $\Delta{M_{bol}}\leq0.05$ mag 
for ages ranging between 10 and 20 Gyr. So we can safely state that the TRGB
luminosity is approximately constant in this age interval. For obtaining a  
variation of the TRGB luminosity of about $\Delta{M_{bol}}\approx0.10$ mag, the
age of the stellar population has to be largely decreased. 
By computing additional evolutionary models with
higher masses at metallicities Z=0.0001 and Z=0.006 we find,
for instance, that at Z=0.0001
a reduction of about $\Delta{M_{bol}}\approx0.10$ mag in the TRGB 
luminosity is obtained for an age around 2.8Gyr, while at Z=0.006 the same variation 
can be found decreasing the age till about 2.2Gyr.
This occurrence is related  to the behavior of the TRGB luminosity through the
Red Giant Phase Transition (Sweigart, Greggio \& Renzini 1989,1990), i.e.
the luminosity of TRGB is almost constant until the evolving mass on the RGB is lower
than the critical mass $M_{HeF}$ which marks the transition between full degenerate
stellar structures which ignite the He burning by a recurrent series of flashes inside
the He core and the ones which ignite quietly the He burning inside a mild degenerate 
core.

In Table 1, the TRGB luminosity and its bolometric magnitude
(obtained adopting for the Sun $M_{bol}=4.75$ mag) are reported for each
adopted metallicity by assuming, as discussed above, an age equal to 15Gyr.
By considering $\rm [M/H]=log(M/H)_{star}-log(M/H)_{\odot}\approx
log(Z)+1.65$, from the data displayed in Table 1 the
following relation is obtained:

$${M^{tip}_{bol}}=-3.949 - 0.178\cdot[M/H] + 0.008\cdot{[M/H]}^2 
\,\,\,\,\,\,\,\,\,\,(1)$$

\noindent
for $-2.35\le[M/H]\le-0.57$, with $\sigma=0.002$.

The enhancement of the $\alpha$ elements observed in galactic field halo and GCs stars
(see, e.g. the review by Wheeler, Sneden \& Truran 1989) is automatically taken into 
account by equation 1 when considering the global metallicity [M/H].
In fact, as already demonstrated by Salaris, Chieffi \& Straniero (1993), 
Paper I, Salaris, Degl'Innocenti \& Weiss (1997), $\alpha$-enhanced theoretical models 
are well reproduced by solar scaled ones with the same global metallicity.
For fixed values of [$\alpha$/Fe]$\ge$0 and [Fe/H] the global metallicity [M/H]
is given by (see Salaris et al. 1993):

$$\rm [M/H]\approx[Fe/H] + log(0.638\cdot$f$+0.362)\,\,\,\,\,\,\,\,\,\,\,\,
\,\,\,\,\,\,\,\,\,\,\,\,\,\,\,\,\,\,\,\,\,\,\,\,\,\,\,\,\,(2)$$

\noindent
where log($f$)=[$\alpha$/Fe].

Equation 1 depends on the adopted initial Helium content since a variation of Y 
at a fixed metallicity changes the TRGB luminosity because the change of the He core 
mass
at the He flash.
As it is well known, the knowledge of the "correct" He enrichment ratio is a 
longstanding
problem, and an exhaustive discussion about this point is beyond the scope of the
present work. However, we want to recall that regardless of the adopted
$\Delta{Y}/\Delta{Z}$ law and for reasonable choices about it, i.e. 
$1<\Delta{Y}/\Delta{Z}<5$ (the reader interested
to recent discussions on this parameter and to the
discrepancy between theoretical and observational
values is referred to Peimbert 1993 and Carigi et al. 1995), 
the He abundance that one has to adopt at $Z\le0.006$
is not very much different from the  
cosmological value $Y\approx$0.23; the maximum variation is of about +0.03 at Z=0.006. 
For studying theoretically the influence of a variation of Y on the TRGB luminosity
we have computed RGB models at Z=0.0001 and Z=0.006, adopting different initial He 
contents
around Y=0.23.
We find that
${\partial{M_{bol}^{tip}}}\over{\partial{Y}}$ is $\approx1.21$ at Z=0.0001 and
$\approx0.75$ at Z=0.006. Therefore, for reasonable choices of 
$\Delta{Y}/\Delta{Z}$, $M_{bol}^{tip}$ 
will vary at most by only $\approx+0.02$ mag at Z=0.006.

Since very recently a new set of EOS tables, that constitutes an improvement
in comparison with older tabulations, has been published (OPAL EOS, see Roger, Swenson \& 
Iglesias 1996), we have also tested the influence of this new EOS on the
theoretical determination of the TRGB luminosity. The OPAL EOS has been 
implemented in the evolutionary code as described in Paper II, and TRGB
luminosities have been computed at the same metallicities previously quoted.
The derived values of $M_{bol}^{tip}$ are higher by not more than 0.02 mag in
comparison with the ones derived from relation 1. Therefore we can consider relation 1
as representative of the theoretical canonical evolutionary determination of the TRGB
luminosity (for metallicities $-2.35\le[M/H]\le-0.57$, Y=0.23)
adopting the most updated available input physics.

\section{\bf The RGB Tip Luminosity: comparison with other distance scales}

\tx

For assessing the reliability of our TRGB luminosities
for distance determinations,  
it is important to compare our results with
observations of GCs and of nearby galaxies in which the stellar 
component has been resolved.
In particular it is important to compare the distance moduli 
of the same object obtained by different
distance indicators, verifying their mutual consistency.
In the following we will separately discuss this kind of comparison for galactic
GCs and for resolved galaxies.

\subsection{\bf Globular Clusters}

\tx

In the case of galactic GCs it is possible to compare the distance scale fixed
by ZAHB models, with the one derived from
relation 1.
Here we have adopted the same ZAHB distance scale presented in Paper II
(the reader is referred to this paper for a comparison of our ZAHB models
with observations), based on
updated stellar models completely consistent with the
RGB ones previously presented:

$$\rm M^{zahb}_V= 1.129 + 0.388\cdot[M/H] + 
0.063\cdot[M/H]^2\,\,\,\,\,\,\,\,\,\,\,\,\,(3)$$ 

\noindent
in the same metallicity range as for equation 1, 
with a r.m.s.=0.011 mag. This relation provides the absolute visual magnitude 
$M^{zahb}_V$
for the ZAHB at $\log{T_{eff}}=3.85$ (that is approximately the average temperature of 
the 
RR Lyrae instability strip) as a function of the global amount of heavy elements [M/H].
By adopting the new OPAL EOS not only for RGB models
but also for the ZAHB phase, luminosities brighter by $\approx0.05$ mag are obtained
in comparison with the values derived from relation 3 (see Paper II).

The comparison between the TRGB and the ZAHB distance scales fixed by equations 1 and 3 
has been performed by adopting the TRGB observational data by
Frogel, Persson \& Cohen (1983 - hereinafter FPC83 - and reference therein), who 
provided absolute bolometric magnitudes for many TRGB of galactic GCs. These magnitudes
have been 
obtained empirically by directly integrating the flux from the program stars
via the observed {\sl UBVJHK} photometry
and adopting a RR Lyrae distance scale for the studied clusters. 
In this way we can compare directly the results from evolutionary computations with 
observations, without using transformations from the theoretical
plane to the observational one.

Eleven clusters for which high resolution spectroscopical metallicity estimates and 
a good determination of their ZAHB level were available, have been selected from the
FPC83 sample.
The adopted ZAHB levels come from Paper II for 7 out of the 11 selected clusters
(see paper II for the sources of the photometric data).
For M71, NGC6352, NGC362 and M15 we have used the photometries respectively 
by Hodder et al. (1992), Fullton et al. (1995), Harris (1982) and 
Durrell \& Harris (1993). In the case of M71, NGC6352 and NGC362, which have a red HB, 
the observational ZAHB level has been determined as described in Paper II for 47 Tuc, 
while for M15 we have adopted the same procedure described in Paper II for M68.
For the clusters in common with DA90 we have adopted their reddening estimates, 
for M3 and M79 the values given by FPC83 have been used, while in the case
of M71, NGC6352 and M68 we have used the values given by, respectively, Hodder et al. 
(1992), Fullton et al (1995) and Walker (1994).

In performing the comparison between RR Lyrae and TRGB distance scales 
we have used our RR Lyrae ZAHB distance scale (homogeneous with the TRGB one)
determined by means of equation 3. 
The observational $M_{bol}^{tip}$ values given by FPC83 were corrected for 
taking  properly into account the differences with respect to
our RR Lyrae distance scale and
our adopted observational ZAHB luminosities.
Moreover, following the prescriptions given by Frogel, Persson \& Cohen (1978)
the values of $M_{bol}^{tip}$ have been also corrected (when necessary)
for the difference between the
reddenings adopted in this paper and the ones used by FPC83. 
The values of
$[Fe/H]$ and $[{\alpha}/Fe]$ obtained by means of spectroscopic analysis, 
the  global  metallicity [M/H] as derived from the  spectroscopic
determinations of $[Fe/H]$ and [$\alpha$/Fe] and using the relation 2), reddening, 
the distance modulus (reddening corrected) and $M_{bol}^{tip}$  are displayed in Table 2 
for the sample of globular clusters we have selected.
\table{2}{S}{\bf Table 2. \rm Data for the sample of selected galactic
globular clusters.} 
{\halign{%
\rm#\hfil& \hskip7pt\hfil\rm#\hfil &\hskip7pt\hfil\rm#\hfil &\hskip7pt\hfil\rm#\hfil &\hskip7pt\hfil\rm#\hfil
&\hskip7pt\hfil\rm#\hfil&\hskip7pt\hfil\rm\hfil#&\hskip7pt\hfil\rm\hfil#\cr
Cluster& [Fe/H] & $[\alpha/Fe]$ & [M/H] &E(B-V)&$(m-M)_{o}$ & $\rm {M}^{tip}_{bol}$ \cr 
\noalign{\vskip 10pt}
M71    & -0.80 & 0.27 & -0.61 & 0.28 & 12.90 & -3.56 \cr 
NGC6352& -0.80 & 0.13 & -0.70 & 0.21 & 13.82 & -3.80 \cr
47 Tuc & -0.80 & 0.15 & -0.70 & 0.04 & 13.19 & -3.65 \cr   
NGC362 & -1.20 & 0.23 & -1.04 & 0.06 & 14.50 & -3.28 \cr
M5     & -1.40 & 0.30 & -1.19 & 0.03 & 14.30 & -3.21 \cr
M79    & -1.42 & 0.21 & -1.27 & 0.00 & 15.62 & -3.54 \cr
NGC6752& -1.50 & 0.31 & -1.28 & 0.04 & 13.00 & -3.44 \cr
M3     & -1.49 & 0.26 & -1.31 & 0.00 & 15.03 & -3.37 \cr
NGC6397& -1.88 & 0.25 & -1.70 & 0.18 & 11.83 & -3.20 \cr
M68    & -1.92 & 0.20 & -1.78 & 0.07 & 14.87 & -3.34 \cr
M15    & -2.30 & 0.30 & -2.09 & 0.10 & 15.03 & -3.37 \cr}}

In Figure 1 we have plotted the $M_{bol}^{tip}$ values  against the global heavy elements
abundance for the selected clusters and also the theoretical relation
for the TRGB luminosity (equation 1).
The vertical error bar ($\pm0.1$ mag) for the observational points 
represents an average error on the distance modulus obtained from relation 3 (see Paper II), while
the error on the spectroscopic determination of [M/H] is typically of about
0.15 dex (see Paper II).

In the case of M5 we have considered an
error bar for the observational $M_{bol}^{tip}$ equal to $-0.3,+0.1$ mag.
This choice is consistent with the comments for this cluster in Table 29 of FPC83.
In fact FPC83 have emphasized the finding, in the central region of M5,
of three stars that result to be up to 0.3 mag brighter than the brightest RGB star
included in their analysis. This correction by 0.3 mag has been also adopted by DA90.

Let us mention again that the theoretical values for the TRGB
luminosity have been converted to $M_{bol}$ values by assuming
the same value of the solar bolometric magnitude as adopted by FPC83 
(i.e.,$M_{bol,\odot}=4.75$ mag).
\figure{1}{S}{100mm}{\bf Figure 1. \rm 
The bolometric magnitude of the brightest observed red giant as a function of the global
metallicity, for the sample of clusters selected from the FPC83 database. 
The solid line shows the theoretical expectation for the bolometric magnitude of the RGB
tip for an age of 15Gyrs. The dashed lines represent the same theoretical relation but
shifted at step of 0.1 mag (see text).}

From a first inspection of Figure 1 it is evident that the TRGB
observational points are located at lower luminosities with respect to the
theoretical relation, with an average difference of 0.20-0.25 mag.
This means that TRGB and ZAHB distance scales agree at least at the level
of $\approx$0.2 mag.
However, it is worth bearing in mind 
that the observational determinations of GCs $M_{bol}^{tip}$ 
as given by FPC83, provide only a lower limit
to the \lq{real}\rq\ maximum luminosity of the TRGB of GCs (see
also the discussion in Castellani, Degl'Innocenti \& Luridiana 1993
and Madore \& Freedman 1995),
since they have observed only a few stars in the upper part of the
RGB of the studied clusters.

To go deeper into the investigation, we have studied if the observed distribution
of the $M_{bol}^{tip}$ values is compatible with the theoretical
models, taking into account the statistical uncertainties due to the
small sample of stars observed.
For this aim, by using our evolutionary models, we have computed the time spent by a 
giant star in a given luminosity interval below the TRGB. The following relation has been 
obtained:
$$t\approx9.67\cdot({\Delta{\log}L}) - 0.43\cdot({\Delta{\log}L})^2 + 
9.82\cdot({\Delta{\log}L})^3 \,\,\,(4)$$

\noindent
where ${\Delta{\log}L}= \log({L/L_{\odot}})^{tip} - \log({L/L_{\odot}})$ and the time 
is in millions of years. This relation is largely independent of the metallicity.

From the ratio of evolutionary times in magnitude intervals $\Delta{M_{bol}}$=0.1 mag,
it is easy to estimate the expected distribution of $\bar{N}$ stars along the last
two bolometric magnitudes as given by $N_i=\bar{N}\cdot{P_i}$ where $P_i$ 
is the probability to find one single star in the chosen interval
(and it is equal to the ratio between the time spent in the selected $M_{bol}$ interval
and the total time spent in the two last bolometric magnitudes below the TRGB).
Adopting from FPC83 $\bar{N}=20$ as a typical value for 
clusters in our sample and using a binomial distribution, the probability to
find at least one star in a chosen luminosity interval below the tip is equal to:

$$P_1=\sum_{n=1}^{20}{{20!}\over{n!\cdot{(20-n)!}}}\cdot{{P_i}^n}\cdot{(1-P_i)}^{20-n} 
\,\,\,\,\,\,\,\,\,\,\,\,\,\,\,\,\,\,\,\,\,\,\,\,\,\,\,\,(5)$$

By using this relation and equation (4), one obtains that it exists a probability
of 56\% of finding at least one star within 0.2 mag below the TRGB,
71\% within 0.3 mag and, 81\% within 0.4 mag below the tip. These values are
in agreement with the data shown in figure 1.
Therefore, in the case of galactic GCs, the TRGB and ZAHB distance scales given by 
relations 1 and 3 agree within the statistical uncertainty due to the small 
stellar sample considered in the observational data.
It is worth noting that if we had used the ZAHB and RGB models computed by adopting the OPAL EOS,
the agreement between the two distance scales would be not degraded, since the variations are
small, especially if compared with the typical observational error. In fact
the points in figure 1 would
be located at luminosities higher by $\approx0.05$ mag (due to the more luminous
ZAHB and therefore higher distance moduli) and the solid line would be lowered by
only $\approx0.01$ mag (due to the slightly lower TRGB luminosity).
\table{3}{S}{\bf Table 3. \rm Dereddened (V-I)
color of the RGB - taken at $M_I=-3.5$ mag - and metallicity
for a sample of globular clusters. The distance moduli of these clusters have been 
obtained adopting our ZAHB distance scale and are displayed in Table 2.} 
{\halign{%
\rm#\hfil&\hskip10pt\hfil\rm#\hfil&\hskip10pt\hfil\rm\hfil#\cr
Cluster & [M/H] &  $(V-I)_{0,-3.5}$ \cr 
\noalign{\vskip 10pt}
M15     &  -2.09  &  1.235~~~  \cr 
NGC6397 &  -1.70  &  1.272~~~  \cr
NGC6752 &  -1.28  &  1.401~~~  \cr   
M5      &  -1.19  &  1.450~~~  \cr
NGC362  &  -1.04  &  1.489~~~  \cr
47Tuc   &  -0.70  &  1.860~~~  \cr}}

DA90 performed a similar comparison for a sample of 8 GCs,
but they used the HB distance scale by Lee, Demarque \& Zinn
(1990) and the theoretical RGB models by Sweigart \& Gross (1978), computed with old input physics. 
Their results were different from ours, since they obtained that the observed values of 
$M_{bol}^{tip}$ were brighter than the theoretical 
prescription by $\approx$0.10 mag. Moreover,
the statistical correction due to the small sample of RGB stars considered would increase 
this discrepancy.
On the contrary, we find that the observational values of 
$M_{bol}^{tip}$ are less bright than the theoretical counterpart
by an amount that is in agreement with the statistical
expectation previously described. This is an important confirmation of the consistency between
RGB and HB evolutionary models, and of the reliability of the input physics used in computing the 
stellar models. In particular, 
it can be also deduced that there cannot be a significant 'missing physics'
that could modify appreciably the degenerate He core masses at the TRGB, which fix the TRGB
and the ZAHB luminosities.

\subsection{\bf Resolved galaxies}

\tx

Another test for checking the consistency of the TRGB luminosity with independent
distance indicators can be performed by considering resolved galaxies, for which
the distance modulus can be derived by using at the same time TRGB, RR Lyrae
stars and Cepheids observations.

LMF93 suggested an iterative procedure for determining the distance of a galaxy 
from observations in the VI Johnson-Cousins bands, by adopting the TRGB 
as distance indicator. Such procedure can be summarized as follows
(see LFM93 for more details):
\medskip
\noindent
i) fixing preliminarly the distance modulus;
\smallskip
\noindent
ii) with the fixed distance modulus determining the metallicity by
measuring the dereddened color at $M_{I}=-3.5$ mag and using the relation:
$[Fe/H]=-12.64+12.6\cdot[(V-I)_{0,-3.5}]-3.3\cdot[(V-I)_{0,-3.5}]^{2}$;
\smallskip
\noindent
iii) obtaining the distance modulus from the observed I magnitude of
the TRGB (corrected for the interstellar extinction) by adopting empirical
relations (from DA90)
for both the TRGB bolometric magnitude as a function of metallicity and the 
bolometric correction to the I magnitude;
\smallskip
\noindent
iv) iterating the previous steps until convergency is obtained between
the distance modulus at step (i) and the one obtained after step (iii).
\medskip
\noindent

Therefore in order to estimate the distance modulus of a galaxy by adopting 
the TRGB method, from a theoretical point of view, one needs
relations providing the bolometric correction in the I band and the bolometric
magnitude of the TRGB. 
As far as the bolometric magnitude of the TRGB is concerned, 
we have already provided a relation which gives the value of $M^{tip}_{bol}$ as a
function of the global
amount of heavy elements (equation 1). Following LFM93,
an empirical $BC_{I}-(V-I)_{o}$ relation for RGB stars has been
taken from DA90. In that paper the authors give:

$$ BC_I= 0.881- 0.243\cdot(V-I)_0 \,\,\,\,\,\,\,\,\,\,\,\,\,\,\,\,\,\,
\,\,\,\,\,\,\,\,\,\,\,\,\,\,\,\,\,\,\,\,\,\,\,\,\,\,\,\,\,\,\,\,\,(6)$$

\noindent
independent of the metallicity,
with a dispersion of 0.057 mag (see their Fig. 14).
In deriving this relation DA90 have considered 
$VI$ photometry (from their observations and from Lloyd Evans 1983)
and the corresponding $M_{bol}$ values (from FPC83)
for stars along the  RGBs of 47Tuc, NGC362, NGC1851, M5,
NGC6752, $\omega$ Cen, NGC6397 and  M15, spanning a wide range of
$(V-I)$ and [M/H] values.
The colors of the stars have been corrected for the reddening of the clusters 
adopting the $E(V-I)/E(B-V)$ relation of Dean, Warren
\& Cousins (1978), 
while the $I$ magnitudes and the $M_{bol}$ values have been adjusted to the
theoretical RR Lyrae distance scale given by Lee, Demarque \& Zinn (1990). 
The difference between $M_{I}$ and $M_{bol}$ gives the requested bolometric 
correction to the I magnitude. It is important to note that, as it is evident,
once the $I$ and $M_{bol}$ data are adjusted to the same distance scale, the
derived $BC_{I}$ is independent on which distance scale is assumed.

Since the bolometric magnitude of the TRGB depends on the metallicity,
and for very distant objects it is not possible to perform accurate high resolution 
spectroscopical determinations of [M/H] as in the case of galactic GCs, 
it is important to obtain a relation providing the heavy elements abundance as 
a function of some observable quantity.
It is well known that the RGB location in the HR diagram ranks with the
metallicity in the sense that the RGB becomes redder and less steep
with increasing [M/H] value.
Therefore one can relate the dereddened color of the RGB at a fixed
absolute magnitude, 
for instance at $M_I=-3.5$ mag ($(V-I)_{0,-3.5}$), following the choice
made by LFM93, with the global amount of heavy elements.
We have selected from the database of DA90, a sample of clusters for
which high resolution, spectroscopical
determinations of [M/H] are available (see Paper I) and have performed  a cubic
regression of the cluster metallicity  versus $(V-I)_{0,-3.5}$ (see
data in Table 3 and figure 2), imposing that the $(V-I)_{0,-3.5}$ values have to be 
monotonously increasing for increasing metallicity, 
obtaining:

$$[M/H]=-45.16 + 73.71\cdot[(V-I)_{0,-3.5}]  {\hskip 2.5truecm}$$
$$\hskip 0.5truecm - 40.91\cdot[(V-I)_{0,-3.5}]^2 + 7.59\cdot[(V-I)_{0,-3.5}]^3 
\,\,\,\,\,(7)$$

\noindent
with a dispersion of 0.04 dex.
The errors displayed in figure 2 correspond to 0.15 dex in [M/H] and
to about 0.04 mag in $(V-I)_{0,-3.5}$, that is the average variation of $(V-I)_{0,-3.5}$ 
due to an uncertainty of about 0.10 mag in the adopted clusters distance moduli.
\figure{2}{S}{100mm}{\bf Figure 2. \rm Metallicity calibration: [M/H] values from Paper 
I, together with their associated uncertainties, are plotted against $(V-I)_{0,-3.5}$.
The solid line shows a cubic fit to these points (see text).}

By adopting relations 1, 6 and 7 it is possible to apply
the TRGB method for distance determinations - as described by LFM93 - 
to nearby resolved galaxies, according to the relation:

$$(m-M)_{I}=I_{TRGB}+BC_{I}-M^{tip}_{bol} \,\,\,\,\,\,\,\,\,\,\,\,\,\,\,\,\,\,\,\,
\,\,\,\,\,\,\,\,\,\,\,\,\,\,\,\,\,\,\,\,\,\,\,\,\,\,\,\,\,(8)$$

\noindent
An alternative approach for applying the TRGB method to resolved galaxies 
is to adopt theoretical $BC_{I}$ values derived from model atmospheres. 
Courtesy of Dr. F. Castelli, we have been able
to use new color transformations, obtained with an updated version
of the Kurucz's code ATLAS9, which constitute a significant improvement 
in comparison with old 
available evaluations (Castelli 1996, private communication). By adopting
these new transformations, we have transformed our tracks from the theoretical plane into 
the observational one and we have analyzed the behavior of the I (Cousins) magnitude of the 
TRGB {\sl versus} the metallicity. The following relation has been obtained, performing a 
best fit to the data:

$$M_{I}^{tip}=-3.732 + 0.588\cdot[M/H] + 0.193\cdot[M/H]^2 \,\,\,\,\,\,\,\,\,\,\,\,\,(9)$$

\noindent
with a r.m.s.=0.008, spanning the same metallicity range as equation 1.

We have therefore applied the TRGB method by adopting both 
the empirical bolometric corrections given by equation 6, or the results from theoretical
model atmospheres (equation 9). In this way we have checked the consistency with the distances
obtained independently by adopting the Cepheid and our RR Lyrae (equation 3) distance scales and,
we have also tested the agreement between empirical and theoretical $BC_{I}$. 
\figure{3}{S}{100mm}{\bf Figure 3. \rm Comparison of distances for the selected
sample of resolved galaxies, obtained using the TRGB (both empirical and theoretical
$BC_{I}$ values) and the Cepheid distance scale.}
\figure{4}{S}{100mm}{\bf Figure 4. \rm Comparison of distances for the selected
sample of resolved galaxies, obtained using the TRGB (by means of
equation 9), the Cepheid and the RR Lyrae (equation 2) distance scales. The comparison
between the distance moduli obtained using the TRGB method and the RR Lyrae
distance scale but adopting for the RR Lyrae stellar population an average metallicity
equal to [M/H]=-1.5 (see text) is also shown.}
\table{4}{D}{\bf Table 4. \rm Selected parameters (see text) for a sample of resolved 
galaxies.} 
{\halign{%
\rm#\hfil & \hskip6pt\hfil\rm#\hfil &\hskip6pt\hfil\rm#\hfil 
&\hskip6pt\hfil\rm#\hfil&
\hskip6pt\hfil\rm#\hfil &\hskip5pt\hfil\rm#\hfil & \hskip6pt\hfil\rm#\hfil & 
\hskip7pt\hfil\rm#\hfil & \hskip7pt\hfil\rm#\hfil & \hskip7pt\hfil\rm\hfil#\hfil\cr
Galaxy & Type & $E(B-V)$ & $I_{TRGB}$ & $[M/H]$ & $(m-M)_{TRGB}$ & $(m-M)_{Ceph}$ & 
$(m-M)_{RR}$ & $(m-M)_{RR}^{-1.5}$ & $(m-M)_{TRGB}^{theor}$ \cr 
\noalign{\vskip 10pt}
LMC     & SBmII       & 0.10  & 14.60 & -1.06 & 18.59 & 18.50 & 18.40 &       & 18.53\cr
NGC6822 & ImIV-V      & 0.28  & 20.05 & -1.75 & 23.61 & 23.62 &       &       & 23.64\cr
NGC185  & dE3pec      & 0.19  & 20.30 & -1.06 & 24.11 &       & 23.93 & 24.01 & 24.05\cr
NGC147  & dE5         & 0.17  & 20.40 & -0.85 & 24.28 &       & 24.00 & 24.13 & 24.14\cr
IC1613  & ImV         & 0.02  & 20.25 & -1.15 & 24.44 & 24.42 & 24.21 & 24.27 & 24.36\cr
M31     & SbI-II      & 0.08  & 20.55 & -0.81 & 24.58 & 24.44 & 24.31 & 24.45 & 24.47\cr
M33     & Sc(s)II-III & 0.10  & 20.95 & -2.05 & 24.82 & 24.63 & 24.69 & 24.62 & 24.87\cr
WLM     & ImIV-V      & 0.02  & 20.85 & -1.48 & 24.97 & 24.92 &       &       & 24.99\cr
NGC205  & S0/dE5pec   & 0.035 & 20.45 & -0.81 & 24.56 &       & 24.68 & 24.82 & 24.46\cr
Sex A   & ImV         & 0.075 & 21.79 & -1.80 & 25.91 & 25.85 &       &       & 25.80\cr
NGC3109 & SmIV        & 0.04  & 21.55 & -1.48 & 25.61 & 25.50 &       &       & 25.65\cr}}

The observational database is the same one collected by LFM93, 
with the exception of the LMC RR Lyrae distances (see discussion below) and
with the additional data for 
Sextans A taken from Sakai et al. (1996). The $A_I$ exctinction has been treated as
in LFM93 (see also Madore \& Freedman 1991, Lee 1993).

As far it concerns the Cepheid distance scale, recently
a big effort has been
made for accounting for metallicity effects in the pulsational properties 
of Cepheid stars (Bono 1996, private communication)
and for metallicity and/or interstellar reddening effects in the observational
calibrations of the Cepheid distance scale (see Laney \& Stobie 1995,
Di Benedetto 1994 and references therein).
However, an analysis of the problems and of the recent theoretical and observational 
improvements in the calibration of the cosmic distance scale through Cepheid variable stars is
beyond the scope of the present paper; here we have adopted the same
distance scale used by LFM93.

In Table 4 we report the distance modulus determinations as obtained with the 
three different 
methods. The various columns provide the following data: (1) the name of the object;
(2) the morphological type (as given by Sandage \& Tammann 1987); (3) the reddening; 
(4) the observed I magnitude of the TRGB;
(5) the mean RGB metallicity, as obtained adopting relation 7;
(6) the distance modulus estimated by using the TRGB method; (7) the intrinsic Cepheid
distance; (8) the distance obtained by using the RR Lyrae luminosity; 
(9) as in column (8) but for an average metallicity of the RR Lyrae population
[M/H]=-1.5 (see below); (10) the distance modulus obtained applying equation 9.

The TRGB luminosities determined for this sample of galaxies 
are based on observations of a large number of RGB stars. According to Madore \& Freedman (1995),
for avoiding an underestimation of the $I$ TRGB magnitude due to population effects (as  
discussed previously ) a sample of about 50 - 100 stars 
in the upper magnitude interval has to be observed. In the case of these resolved galaxies the RGB
star sample is large enough to satisfy this requisite.
Therefore we can in principle compare directly observed and predicted TRGB $I$ luminosities,
without taking into account the statistical uncertainty due to the size of the star sample.
The typical observational errors on the estimated distance moduli with the TRGB or the
Cepheids are of the order of $\approx$0.15 mag (see, i.e. LFM93, Sakai et al. 1996).

The RR Lyrae distance moduli displayed in column 8 of Table 4 are obtained considering the
observed RR Lyrae average magnitude, applying the correction for the off ZAHB evolution 
given by equation 5 of Paper II, and then using equation 3; the metallicities considered in 
this procedure are given in column 5 and correspond to the mean RGB metallicities.
In the case of the LMC 
we have considered the average RR Lyrae luminosities for 5 clusters (NGC1466, NGC1835, NGC1841,
NGC2257, Reticulum) with a number of observed RR Lyrae stars larger than 20. The observational
data, the reddenings and the [Fe/H] values for these clusters are taken from Walker (1992).
We have derived the clusters distance moduli assuming 
[$\alpha$/Fe]=0.0 or 0.3, two different corrections due to the LMC geometry (as described
by Walker 1992) and no correction.
After averaging the five distance moduli for each of the six cases 
considered, we have obtained a LMC distance modulus ranging between 
18.38 and 18.44; in Table 4 we have reported 
the value derived without corrections for the LMC geometry, and assuming [$\alpha$/Fe]=0.

The errors associated to the RR Lyrae distance scale derived via equation 3
are probably higher than in the case of the TRGB, at least when the observations 
adopted were performed in the 
$g$ Thuan-Gunn band (as in the case of NGC185, NGC147, IC1613, NGC205).
Infact, in this case it is necessary to transform the $<g>$ values for the
mean RR Lyrae luminosity to $<V>$, before applying the correction for the evolution off the ZAHB and 
equation 3. But after the pioneristic work of Thuan \& Gunn (1976), such transformation
has not been studied in details, so up to date its associated uncertainty 
is unknown. For such a reason, it could be possible that the use of this transformation 
law (in particular we have used the prescription given
by Saha \& Hoessel 1990) introduces an additional uncertainty to the final derived distance modulus. 
Moreover, it is worth bearing in mind that, even if Saha \& Hoessel (1990) 
have shown that the presence of non RR Lyrae pulsators does not significantly
affect the estimate of the mean magnitude obtained from the peak of the magnitude distribution
of {\sl all} pulsating stars, it is possible that in some stellar system such occurrence
could introduce a not negligible uncertainty in the {\sl real} mean RR Lyrae magnitude.

Another important source of uncertainty for the RR Lyrae distance modulus
is related to the metallicity of the RR Lyrae population. The [M/H] value
used for obtaining the distance moduli given in column 8 of Table 4 are derived from
RGB stars, and represent an average metallicity of this stellar population. In principle 
this metallicity could not correspond to the RR Lyrae average metal content, as for
instance for the two "metal-rich RGB" galaxies M31 and NGC205, due to the low probability that
metal-rich RGB stars evolve during their He central burning phase through the RR Lyrae instability strip.
%
%
For roughly estimating the additional uncertainty introduced by the unknown metallicity of the
RR Lyrae population, the distance moduli obtained assuming for the RR Lyrae stars 
an average metallicity equal to [M/H]=-1.5 - adopted as a reasonable estimation of the average
metallicity for the galactic GCs RR Lyrae population - have been also reported 
(with the unique exception of the LMC)
in Table 4 (column 9).

In figure 3, we display the difference between the distance moduli obtained by adopting
the TRGB and the Cepheid distance scale with respect to the TRGB distances
(by using both empirical and theoretical $BC_{I}$ values). 
As it is evident from the figure, the agreement between these three different 
distance scales is remarkably
satisfactory.
In particular, the average difference between the TRGB distance moduli obtained by adopting 
empirical $BC_{I}$ values and the Cepheid distances is equal to only $\approx$+0.08 mag, while 
in the case of theoretical $BC_{I}$ values the same average difference is $\approx$+0.05 mag.

In Figure 4, the differences between TRGB distance scale fixed by equation 9 or
the Cepheid distance scale with respect to the RR Lyrae ones (columns 8 and 9 of Table 4) are
displayed as a function of the RR Lyrae distance moduli. 
When considering for the population of RR Lyrae stars metallicities equal to the [M/H] values
obtained from the RGB color (column 5 in Table 4),
the average difference between RR Lyrae and Cepheid distance moduli is  
of about 0.14 mag (the RR Lyrae distance moduli being lower), 
while between RR Lyrae and TRGB 
is of about 0.12 mag (again the RR Lyrae distance moduli being lower).
If we don't consider NGC205, the average difference 
between RR Lyrae and TRGB distance moduli for the remaining six galaxies results to be
of about 0.18 mag, with a much smaller scatter, the RR Lyrae distance moduli being systematically
lower.
By adopting the OPAL EOS in the evolutionary computations, the distance moduli 
obtained using the RR Lyrae distance scale should be increased by
about 0.05 mag (whereas, as discussed in section 2 the TRGB luminosity is almost 
unaffected by adopting the OPAL EOS in the computations).
In this case the difference between the RR Lyrae and the Cepheid distance scales
is reduced to $\approx$0.10 mag, while the difference with respect to the TRGB distance scale 
is of about 0.13 mag (not considering NGC205). 

If we consider the distance moduli obtained assuming [M/H]=-1.5 for RR Lyrae stellar populations
(and the LMC distance modulus given in column 8 of Table 4) the average difference between 
RR Lyrae and TRGB distance moduli (still neglecting NGC205) is equal to $\approx$0.09 mag 
(0.08 mag if for M33 we consider its metal poor RGB metallicity),
reduced to $\approx$0.04 mag when taking into account the effect of OPAL EOS on the ZAHB models.

In particular, when comparing the LMC RR Lyrae distance modulus (that probably is
more accurate than for the other galaxies and it is less affected from the uncertainties on the real RR Lyrae
metallicity) with the TRGB one, we find a difference
of 0.13 mag (reduced to $\approx$ 0.08 mag if the OPAL EOS is used in the stellar models), 
well within the errors that are typically of the order of 0.15 mag for the TRGB and
at least of 0.10 mag for the RR Lyrae distances.

Therefore from the data presented in this section and in section 3.1 
it comes out that, within the observational uncertainties and in the limit of the small
sample of GCs and resolved galaxies considered, 
it does not appear to exist a clear inconsistency between 
ZAHB and RGB stellar models; the TRGB and RR Lyrae distance scales set by the evolutionary
calculations agree at the level of $\approx$0.1 mag, while the TRGB and Cepheid distance
scales are consistent within less than 0.1 mag.

\section{\bf Summary and conclusions}

\tx

We have performed a study about the self-consistency of theoretical 
RGB and ZAHB stellar models, computed with the most updated input 
physics.
For this aim and for providing an updated distance scale based on the TRGB method,
new theoretical $M^{tip}_{bol}$-[M/H] and $M^{tip}_{I}$-[M/H] relations
have been provided together with a recalibration of the
empirical [M/H]-$(V-I)_{0,-3.5}$ relation 
(based on recent high resolution metallicity determinations 
for galactic GCs). 

The comparison between the TRGB distance scale 
and the RR Lyrae one, adopting galactic GCs observations, has disclosed a good agreement within
the statistical uncertainty due to the size of the observed clusters stellar sample presently
available.

We have presented also a comparison among TRGB, RR Lyrae and Cepheid distance scales, 
employing recent photometric studies of nearby resolved galaxies.
The following results have been obtained:
\medskip\noindent 
i) the distance moduli determined by adopting our $M^{tip}_{I}$-[M/H] relation,
or $M^{tip}_{bol}$-[M/H]
together with the empirical $BC_{I}$ values by DA90, or the Cepheid distance scale by LFM93, agree
on average within less than 0.1 mag; in particular, it is remarkably good the consistency between the
completely theoretical prescription for the $I_{TRGB}$ luminosities given by equation 9 and
the Cepheid distance scale (distance moduli different on average by only 0.05 mag).
\medskip\noindent
ii) the agreement between RR Lyrae and TRGB (or Cepheid) distance scales when considering
the sample of resolved galaxies is at the level of $\approx$0.1 mag, taking into account
the uncertainties on the metallicity of the RR Lyrae populations.  
The data from GCs are consistent with no discrepancy between TRGB and RR Lyrae distances;
\medskip\noindent
iii) the [M/H] values obtained by adopting the recalibrated [M/H]-$(V-I)_{0,-3.5}$ relation 
are higher by not more than 0.15 dex in comparison with the [Fe/H] values derived by LFM93;
\medskip\noindent
iv) the distance moduli derived adopting our theoretical $M^{tip}_{I}$-[M/H] relation, together with 
our estimations of the global metallicities are on average $\approx0.11$ mag higher
than the values obtained by LFM93.
\medskip
\noindent
The satisfactory agreement between the three discussed distances scales  
allows to assess the reliability of the presented theoretical TRGB and ZAHB luminosities.
In particular, the provided theoretical relations for the TRGB luminosity (equations 1 and 9) can
be safely used when adopting the TRGB as a distance indicator for resolved galaxies,
but also for GCs, if a sufficiently populated RGB can be observed.

\section*{Acknowledgments}

\tx 
We thank an anonymous referee for her/his constructive suggestions and pertinent
remarks.
S.C. warmly thanks F. Castelli for many interesting discussions on the Kurucz
model atmospheres and for kindly providing in advance of publication
updated bolometric corrections and color transformations.
We also are pleased to thank G. Bono for interesting
suggestions, discussions and for reading a preliminary draft of this paper
which has improved its readability, M. Groenewegen, L. Piersanti and
S. Schindler for helpful discussions. M.S. warmly
acknowledges the unvaluable human support of 'Caf\'e Ruffert \& Schindler' 
at MPA.

\section*{References}
\bibitem Alexander D. R. \&  Ferguson J. W. 1994, ApJ, 437, 879 
\bibitem Baade W. 1944, ApJ 100, 137 
\bibitem Buzzoni A., Fusi Pecci F., Buonanno R. \& Corsi C. E. 1983, A\&A 128, 94 
\bibitem Carigi L., Colin P., Peimbert M. \& Sarmiento A. 1995, ApJ 445, 98
\bibitem Cassisi S. \& Salaris M. 1997, MNRAS in press
\bibitem Castellani V., Chieffi A. \& Pulone L. 1991, ApJS 76, 911 
\bibitem Castellani V., Chieffi A. \& Straniero O. 1992, ApJS 78, 517
\bibitem Castellani V., Degl'Innocenti S. \& Luridiana V. 1993, A\&A 272, 442
\bibitem Chieffi A., \& Straniero O. 1989, ApJS, 71, 47
\bibitem Da Costa G.S. \& Armandroff T.E. 1990, AJ 100, 162
\bibitem Dean J.F., Warren P.R. \& Cousins A.W.J. 1978, MNRAS 183,569
\bibitem Di Benedetto G. P. 1994, A\&A 285, 819
\bibitem Durrell P.R. \& Harris W.E. 1993, AJ 105, 1420
\bibitem Elson R. A. W. 1996, MNRAS in press
\bibitem Frogel J.A., Persson S.E. \& Cohen J.G. 1978, ApJ 222, 165
\bibitem Frogel J.A., Persson S.E. \& Cohen J.G. 1983, ApJS 53, 713
\bibitem Fullton L. K., Carney B. W., Olszewski E. W., Zinn R., Demarque P. \&
Janes K. A. 1995, AJ 110, 652
\bibitem Green E.M., Demarque P. \& King C.R. 1987, The Revised
Yale Isochrones and luminosity Functions (Yale University Observatory, New Haven)
\bibitem Grevesse, N. 1991, in G. Michaud \& A. Tutukov (eds.) IAU Symp. 145, Evolution of Stars: 
the Photospheric Abundance Connection, (Dordrecht: Kluwer), p. 63      
\bibitem Iglesias C.A., Rogers F.J. \& Wilson B.G. 1992, ApJ 397, 717
\bibitem Itoh N., Mitake S., Iyetomi H. \& Ichimaru S. 1983, ApJ 273, 774
\bibitem Harris W. E. 1982, ApJS 50, 573
\bibitem Hodder P. J. C., Nemec J. M., Richer H. B. \& Fahlman G. G. 1992, AJ 103, 460
\bibitem Kurucz R. L. 1992, in Barbuy B., Renzini A. (eds.), IAU
Symp. n. 149, ``The Stellar Populations of Galaxies'', Kluwer,
Dordrecht, p. 225
\bibitem Laney C. D. \& Stobie R. S. 1995, MNRAS 274, 337
\bibitem Lee M.G. 1993, ApJ 408, 409
\bibitem Lee M.G., Freedman W. \& Madore B.F. 1993, ApJ 417, 553
\bibitem Lee Y.-W., Demarque P. \& Zinn R. 1990, ApJ 350, 155
\bibitem Lloyd Evans T. 1983, S.Afr.Astron.Obs.Circ. 7, 86
\bibitem Madore B. F. \& Freedman W.L. 1991, PASP 103, 933
\bibitem Madore B. F. \& Freedman W.L. 1995, AJ 109, 1645
\bibitem Olive K.A., Skillman E. \& Steigman G. 1996, preprint, astro-ph/9611166
\bibitem Peimbert M. 1993, Rev. Mex. Astr.Astrofis. 27, 9
%
%
\bibitem Rogers F. J. \& Iglesias C. A. 1992, ApJ, 401, 361
\bibitem Rogers F.J., Swenson F.J. \& Iglesias C.A. 1996, ApJ 456, 902
\bibitem Saha A. \& Hoessel J.G. 1990, AJ 99, 97
\bibitem Sakai S., Madore B.F. \& Freedman W.L. 1996, ApJ 461, 713
\bibitem Salaris M. \& Cassisi S. 1996, A\&A 305, 858
\bibitem Salaris M., Chieffi A. \& Straniero O. 1993, ApJ 414, 580
\bibitem Salaris M., Degl'Innocenti S. \& Weiss A. 1997, ApJ in press
\bibitem Sandage A.R. 1971, in D.J.K. O'Connel (ed.), "Nuclei of Galaxies", (Amsterdam:
North-Holland), p. 601
\bibitem Sandage A. R. \& Tammann G. A. 1987 A Revised Shapley-Ames Catalog of
Bright Galaxies (Washington, DC: Carnegie Institution)
\bibitem Soria R. et al. 1996, ApJ 465, 79
\bibitem Straniero O. 1988, A\&AS 76, 157
\bibitem Straniero O. \& Chieffi A. 1991, ApJS 76, 525 
\bibitem Sweigart A. V., Greggio L. \& Renzini A. 1989, ApJS, 69, 911   
\bibitem Sweigart A. V., Greggio L. \& Renzini A. 1990, ApJ, 364, 527  
\bibitem Sweigart A. V. \& Gross P. G. 1978, ApJS 36, 405
\bibitem Thuan T. X. \& Gunn J. E. 1976, PASP 88, 543
\bibitem Walker A. R. 1992, ApJ 390, L81
\bibitem Walker A. R. 1994, AJ 108, 555
%
%
\bibitem Wheeler J.C., Sneden C. \& Truran J. W. 1989, ARAA 252, 179

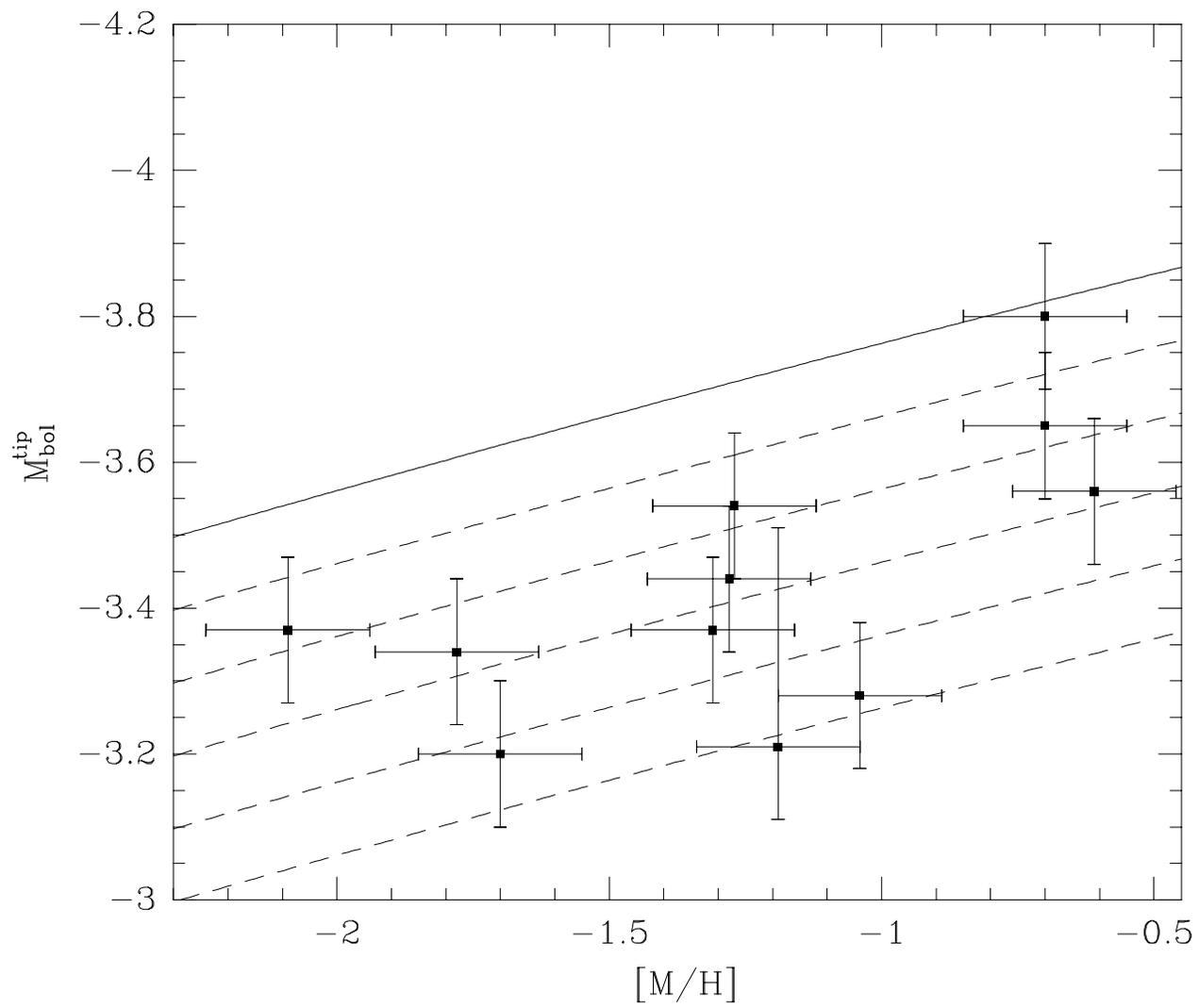
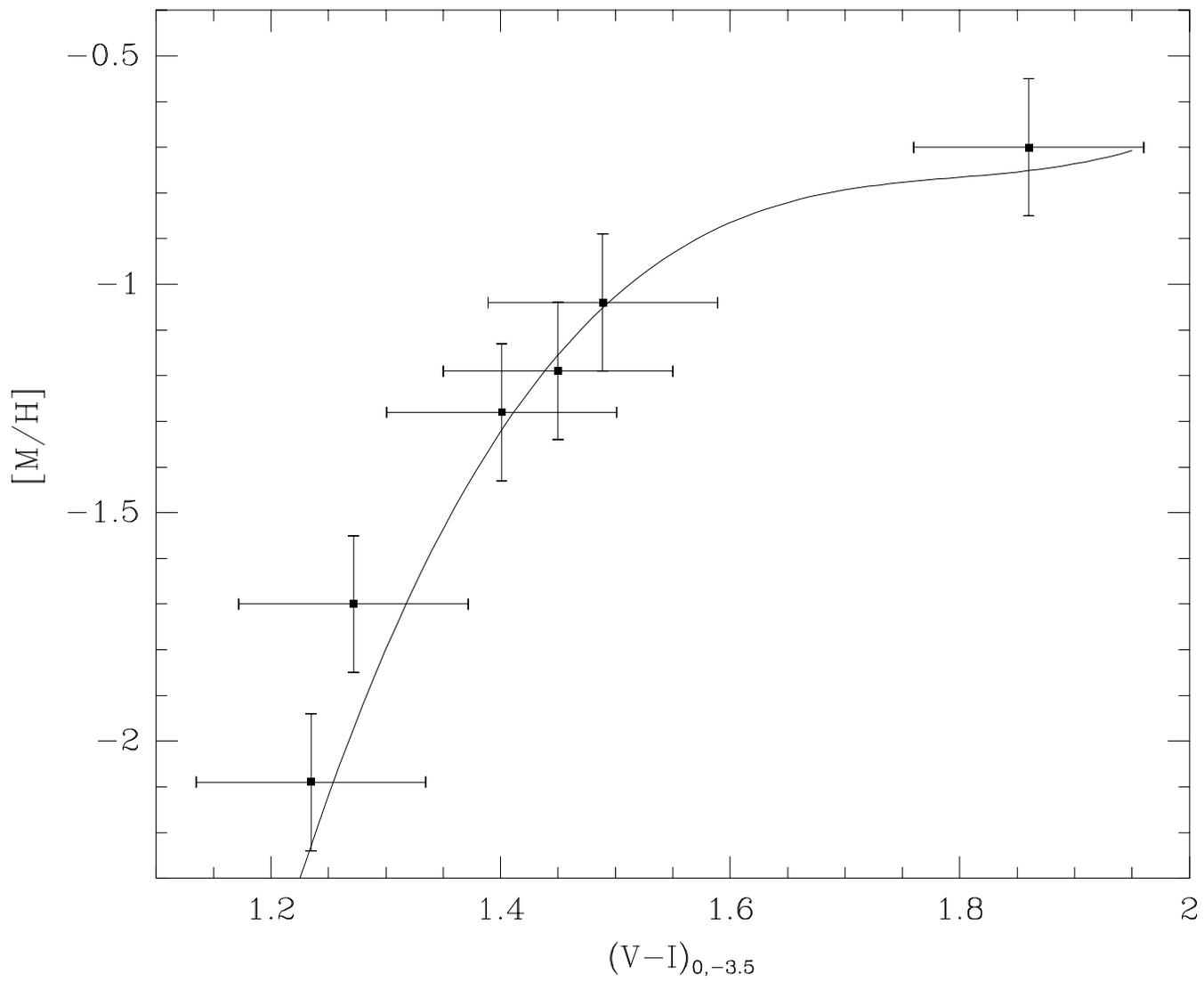
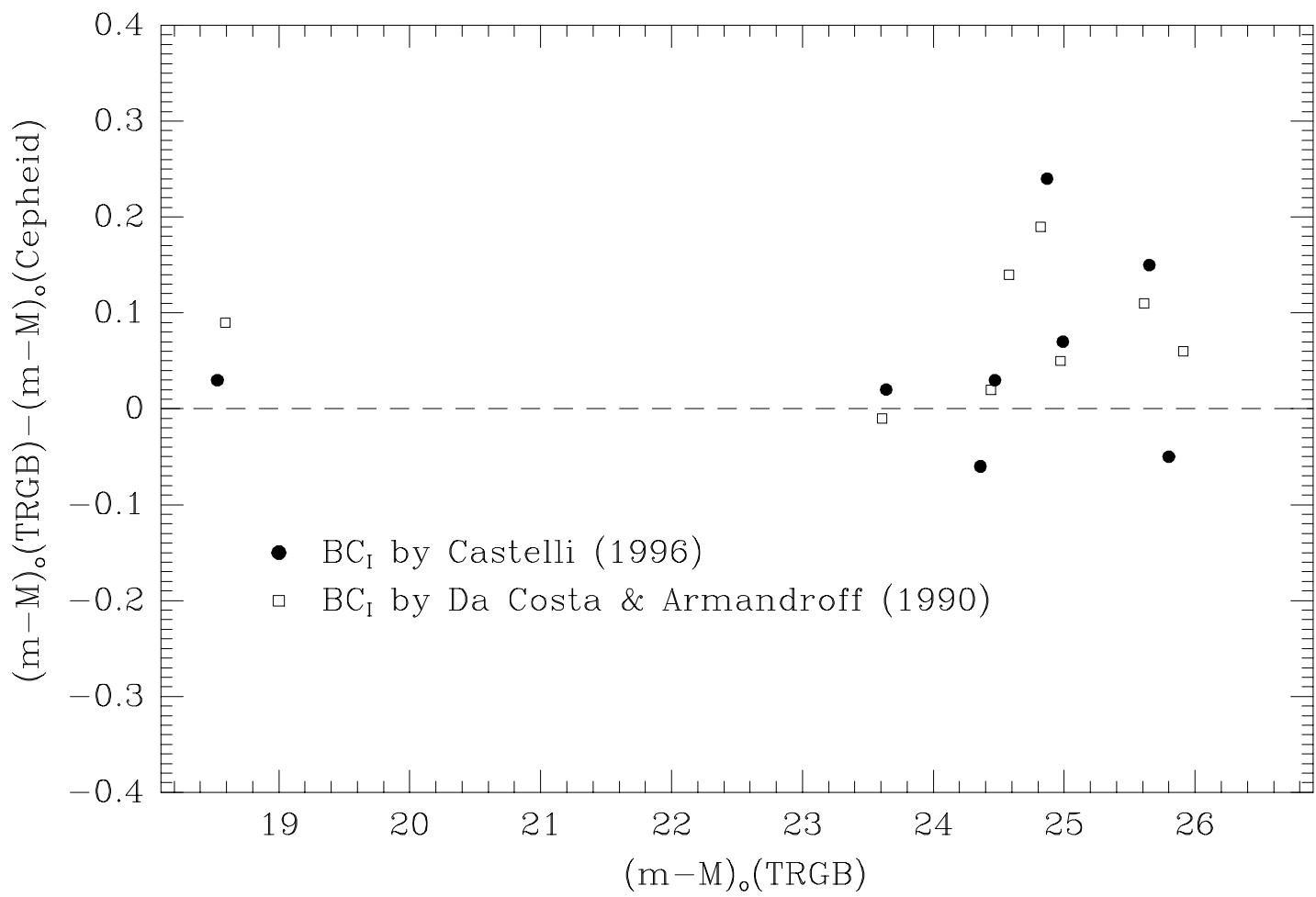
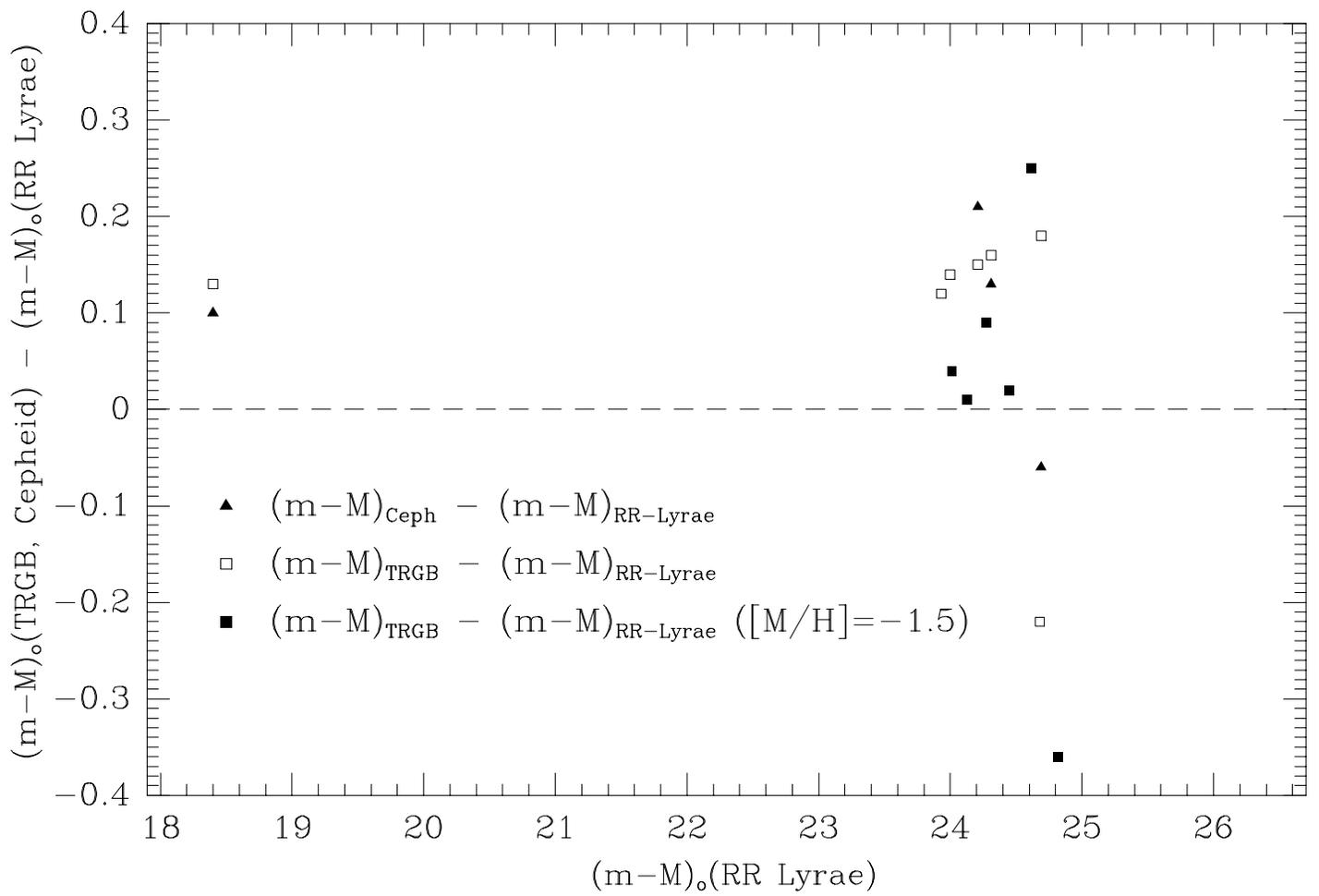
\bye